\documentclass{PoS}

\title{The electromagnetic form factor of the pion in two-flavour lattice QCD}

\ShortTitle{The electromagnetic form factor of the pion in two-flavour lattice
QCD}

\author{\speaker{Bastian B. Brandt}\\
        Institut f\"ur theoretische Physik,\\
        Universit\"at Regensburg, D-93040\\
        E-mail: \email{bastian.brandt@physik.uni-regensburg.de}}

\author{Andreas J\"uttner\\
        School of Physics and Astronomy,\\
        University of Southampton, Southampton, SO14 1BJ, UK}

\author{Hartmut Wittig \\
        PRISMA Cluster of Excellence, \\
        Institut f\"ur Kernphysik and Helmholtz Institut Mainz,\\
        Johannes Gutenberg-Universit\"at Mainz, D-55099}

\abstract{We present the current status of our lattice calculation of the
electromagnetic form factor of the pion with two flavours of non-perturbatively
O(a)-improved Wilson fermions. Using twisted boundary conditions and stochastic
sources we obtain accurate results with a fine momentum resolution near $q^2=0$.
This enables the computation of the charge radius without model dependence. The
ensembles cover various lattice spacings and pion masses, ranging down to 250
MeV. This allows to compare the data to continuum chiral perturbation theory to
NNLO including corrections of finite lattice spacing to perform a simultaneous
chiral and continuum extrapolation. An estimate for the systematic error
resulting from the extrapolation can be obtained by looking at the spread of
results obtained from other functional forms such as polynomials.}

\FullConference{Xth Quark Confinement and the Hadron Spectrum\\
                 8-12 October 2012\\
                 TUM Campus Garching, Munich, Germany}

\begin{document}

\section{Introduction}

Lattice QCD has by now entered the era in which high
precision calculations of physical quantities at the physical point including
dynamical fermionic degrees of freedom become feasible (see for example the
quantities summarised in the FLAG review~\cite{Colangelo:2010et}). This progress
has been possible largely due to substantial algorithmic improvements and
the associated acceleration of the simulations allowing to simulate light
quarks. Despite these successes there are quantities where a
satisfactory control over systematics has not been achieved so far, among
them quantities related to the structure of
hadrons~\cite{Renner:2012yh,Lin:2012ev}. For the pion electromagnetic form
factor the dominant systematic uncertainty is due to the chiral extrapolation.
The form factor has been worked out in chiral 
perturbation theory~\cite{Gasser:1984gg,Gasser:1984ux} ($\chi$PT) to
next-to-next-to leading order~\cite{Bijnens:1998fm,Bijnens:2002hp} (NNLO) which
can be used to guide the chiral extrapolation. However, the pion mass range for
which $\chi$PT to a given order is applicable is not known {\it a priori}.

In this proceedings article we present the status of our calculation of the
pion form factor $f_{\pi\pi}(Q^2)$ in the region of small space-like momentum
transfers $Q^2=-q^2$ in Lattice QCD with two dynamical quark flavours. In this
regime the form factor encodes information about the electromagnetic charge
radius $\left<r_\pi^2\right>$ of the pion,
\begin{equation}
 \label{eq:chrad}
       f_{\pi\pi}(q^2)=1-\frac{1}{6}\langle r_\pi^2\rangle
       q^2+\mathcal{O}(q^4), \qquad
       \langle r^2_\pi\rangle = 6\left.
       \frac{\displaystyle f_{\pi\pi}(q^2)}{\displaystyle q^2}\right|_{q^2=0}.
\end{equation}
Determinations of the charge radius usually suffer from an intrinsic model
dependence since the slope of the form factor is obtained from fits over a
large $Q^2$-interval assuming a particular model for the $Q^2$-dependence such
as pole dominance. Here we use suitably tuned partially twisted
boundary
conditions~\cite{Bedaque:2004kc,Sachrajda:2004mi,deDivitiis:2004kq,Flynn:2005in,
Boyle:2007wg} to compute data points in the immediate vicinity of $Q^2=0$ which
enables the model-independent extraction of the charge radius. The data is
then compared to $\chi$PT which is used to extrapolate the charge radius to the
physical point. The continuum extrapolation is done by introducing terms that
model the cutoff dependence into the chiral extrapolation. The systematic
uncertainty which is introduced by relying on $\chi$PT for our range of pion
masses is taken into account by checking the results from $\chi$PT with results
from a polynomial extrapolation. Parts of the results were already reported
in~\cite{Brandt:2010ed,Brandt:2011sj,Brandt:2011jk,Brandt:2012zza}.

\section{Extraction of the form factor}

The electromagnetic form factor in two-flavour QCD is defined by
\begin{equation}
\label{eq:fpipi}
   \left\langle\pi^+(\vec{p}_f)|
   \frac{2}{3}\bar{u}\gamma_\mu u
  -\frac{1}{3}\bar{d}\gamma_\mu d
   |\pi^+(\vec{p}_i)\right\rangle = (p_f+p_i)_\mu\,f_{\pi\pi}(q^2)\,.
\end{equation}
In lattice QCD the matrix element can be extracted from suitable ratios
of 2- and 3-point functions, in particular we use the ratio~\cite{Boyle:2007wg}
\begin{equation}
\label{eq:ratio}
   R_1(t,\vec{p}_i,\vec{p}_f) = \:Z_V^{\rm eff} \:\: 4 \:
   \sqrt{E(\vec{p}_i)\:E(\vec{p}_f)}
   \:\:\sqrt{\frac{C_{3}(t,\vec{p}_i,\vec{p}_f)\:C_{3}(t,\vec{p}_f,\vec{p}_i)}{
   C_2(T/2,\vec{p}_i)\:C_2(T/2,\vec{p}_f)}},
\end{equation}
where
\begin{equation}
\label{eq:zeff}
   Z_V^{\rm eff} = \frac{\displaystyle C_2(T/2,\vec{0})}
   {\displaystyle 2\:C_3(t,\vec{0},\vec{0})}
\end{equation}
ensures the physical normalisation $f_{\pi\pi}(0)=1$ and
\begin{equation}
\label{eq:correlators}
   \begin{array}{rcl}
      \displaystyle C_2(t,\vec{p}) & = & \displaystyle
      \sum_{\vec{x}_i,\vec{x}}e^{i\vec{p}\cdot(\vec{x}-\vec{x}_i)} \langle
      \:P(t,\vec{x})\: P(0,\vec{x}_i)\:\rangle \,, \vspace*{2mm} \\
      \displaystyle C_3(t,\vec{p}_i,\vec{p}_f) &=& \displaystyle 
      \sum_{\vec{x}_i,\vec{x}_f,\vec{x}} e^{i\vec{p}_f\cdot(\vec{x}_f-\vec{x})}
      e^{i\vec{p}_i\cdot(\vec{x}-\vec{x}_i)} \: \langle\, P(t_f,\vec{x}_f)\:
      V^{I}_0(t,\vec{x})\:P(0,\vec{x}_i)\,\rangle \;.
  \end{array}
\end{equation}
Here $P(x)$ and $V^I_\mu(x)$ are the pseudoscalar density and the
$\mathcal{O}(a)$-improved vector current, respectively, and $E(\vec{p})$ is the
energy of the pion with momentum $\vec{p}$. Up to corrections from excited
states the ratio in eq.~(\ref{eq:ratio}) is constant in Euclidean time and gives
the desired matrix element of the vector current from eq.~(\ref{eq:fpipi}). In
practice, the factor $Z_V^{\rm eff}$ is obtained prior to the extraction of the
matrix element via eq.~(\ref{eq:zeff}).

\section{Simulation setup}

\begin{table}[t]
\centering
\vspace*{-3mm}
\begin{tabular}{ccccccc}
\hline
\hline
$\beta$ & $r_0/a$ & lattice & $m_\pi$ [MeV] & $m_\pi\:L$ & Labels & Statistic
\\
\hline
5.20 & 6.15 (6) & $64\times32^3$ & 455, 351, 299 & 6.0 -- 4.0 & A3 -- A5 &
$\mathcal{O}(100)$ \\
\hline
5.30 & 7.26 (7) & $64\times32^3$ & 552, 412 & 6.2, 4.7 & E4, E5 &
$\mathcal{O}(100)$ \\
 & & $96\times48^3$ & 295, 254 & 5.0, 4.2 & F6, F7 & $\mathcal{O}(250)$ \\
\hline
5.50 & 10.00 (11) & $96\times48^3$ & 625, 534, 424 & 7.7 -- 5.3 & N3 -- N5 &
$\mathcal{O}(100)$ \\
\hline
\hline
\end{tabular}
\caption{Compilation of simulation parameters.}
\label{tab:parameters}
\end{table}

The calculations are based on the CLS ensembles that have been generated with
two degenerate flavours of non-perturbatively $\mathcal{O}(a)$-improved Wilson
fermions using the deflation accelerated DD-HMC~\cite{Luscher:2007es} and
MP-HMC~\cite{Marinkovic:2010eg} algorithms. The bare parameters and some of the
basic properties of the lattices are given in table~\ref{tab:parameters}. We
express dimensionful quantities in units of the Sommer scale
$r_0$~\cite{Sommer:1993ce}, which was recently determined on the CLS
ensembles~\cite{Leder:2010kz,Fritzsch:2012wq} with a physical value of
$r_0=0.503(10)$~fm~\cite{Fritzsch:2012wq}.

All 2- and 3-point functions were computed using two hits of stochastic
$Z_2\times Z_2$ wall
sources~\cite{Foster:1998vw,McNeile:2006bz,Boyle:2008rh,Endress:2011jc} on two
different timeslices separated in time by $T/2$. The only exception is
ensemble A5 where we have used four source positions separated by $T/4$.
We are primarily interested in the region of small momentum transfers which
is made accessible by partially twisted boundary conditions at vanishing Fourier
momentum. For each ensemble we have used five different twist angles in the
$x$-direction, that are tuned to achieve a high resolution in the low $Q^2$
region. In addition to the form factor we have also computed the
renormalised PCAC quark mass $\hat{m}$ and pion decay constant $F_\pi$. All
quantities are fully $\mathcal{O}(a)$-improved, and we adopt the renormalisation
constants
from~\cite{DellaMorte:2005kg,DellaMorte:2005se,DellaMorte:2008xb,
Fritzsch:2010aw,Fritzsch:2012wq}. The results are listed in
table~\ref{tab:basicmeasurements}. Note that our ensembles cover a large range
of quark masses and lattice spacings, always satisfying $m_\pi L\geq4$.

Statistical errors have been estimated using the bootstrap procedure with 1000
bins. At present all fits for the extraction of the masses and matrix elements
are uncorrelated.

\section{Results for form factor and charge radius}

\begin{figure}[t]
\begin{center}
 \includegraphics[width=0.7\textwidth]{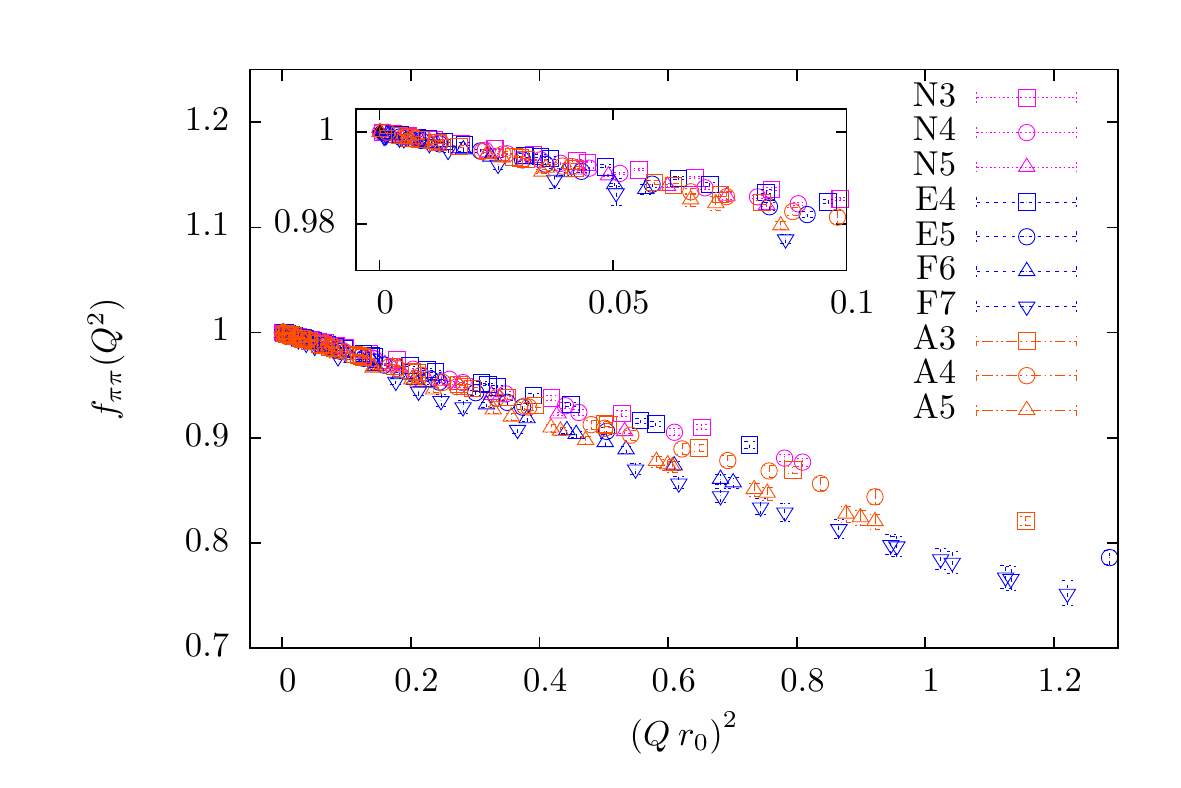}
\vspace*{-3mm}
\caption{Results for the form factor on all ensembles. The inset shows a zoom
in the region of low momentum transfer.}
\label{fig:formfactor}
\end{center}
\end{figure}

The full set of results for the form factor is shown in
figure~\ref{fig:formfactor}. The plot demonstrates that partially twisted
boundary conditions are a powerful tool to explore the small momentum transfer
region of form factors. In particular, the high density of points in the low
$Q^2$ region enables the accurate and model independent extraction of the
charge radius, as discussed below.

The data is still subject to effects stemming from the finite box size, even
though they are expected to be small, since $m_\pi L\geq4$. We estimate their
size in $\chi$PT. For the pion mass and the decay constant the relevant
expressions have been derived in~\cite{Colangelo:2005gd}. For the form
factor with partially twisted boundary conditions, finite size effects are only
known for the case where one of the twists is zero~\cite{Jiang:2006gna}, i.e.
either the initial twist $\theta_i=0$ or the final twist $\theta_f=0$, and in
the Breit frame~\cite{Jiang:2008te} where $\theta_i=-\theta_f$. As input for the
$\chi$PT formulae NLO low energy constants (LECs) are needed. To become
independent of external input we apply an iterative procedure: First we estimate
the finite volume effects using the LECs from~\cite{Colangelo:2001df} as input.
We then perform the extraction of the charge radius and the form factor and fit
the data to NNLO $\chi$PT. This yields new LECs that are used for the next
iteration. After two such iteration this procedure converged and we did not see
any residual deviations in the LECs.

\begin{table}
\begin{center}
\vspace*{-3mm}
\begin{tabular}{lcccc}
\hline\hline&&\\[-4mm]
ensemble & $r_0\:m_\pi$ & $r_0\:F_\pi$ & $r_0\:\hat{m}$ & $Z_V^{\rm{eff}}$ \\
\hline\hline&&\\[-4mm]
A3 & 1.161(12) & 0.280( 8) & 0.090(3) & 0.73228( 7) \\
A4 & 0.895(11) & 0.247(12) & 0.052(3) & 0.72885(12) \\
A5 & 0.761(11) & 0.251(14) & 0.040(2) & 0.72731(10) \\
\hline                                                    
E4 & 1.406(16) & 0.287(10) & 0.128(5) & 0.74962( 8) \\
E5 & 1.048(13) & 0.271(11) & 0.078(4) & 0.74461( 8) \\
F6 & 0.752( 8) & 0.254( 8) & 0.041(2) & 0.74119( 4) \\
F7 & 0.646( 7) & 0.237( 8) & 0.029(1) & 0.74030( 5) \\
\hline
N3 & 1.593(18) & 0.329( 7) & 0.188(5) & 0.77162( 3) \\
N4 & 1.360(16) & 0.304( 9) & 0.139(4) & 0.76855( 3) \\
N5 & 1.080(13) & 0.291( 8) & 0.091(3) & 0.76543( 3) \\
\hline\hline
\end{tabular}
\caption{Results for basic quantities.}  
\label{tab:basicmeasurements}
\end{center}
\end{table}

\begin{figure}[t]
\centering
\vspace*{-3mm}
\begin{minipage}{0.48\textwidth}
\centering
 \includegraphics[width=1.1\textwidth]{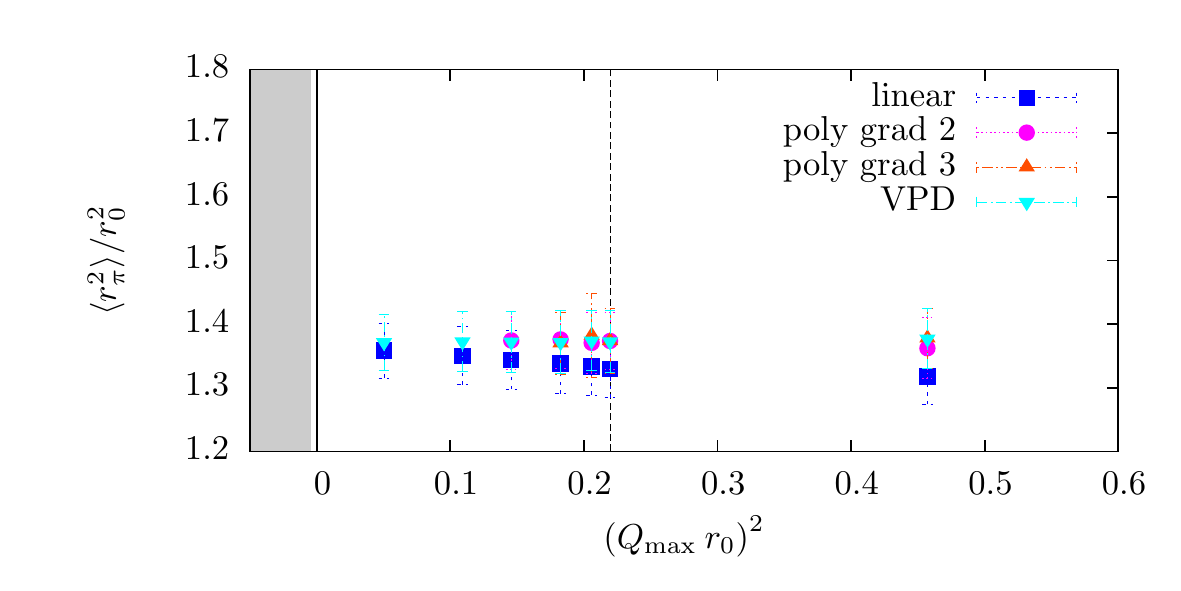}
\end{minipage}
\begin{minipage}{0.48\textwidth}
\centering
 \includegraphics[width=1.1\textwidth]{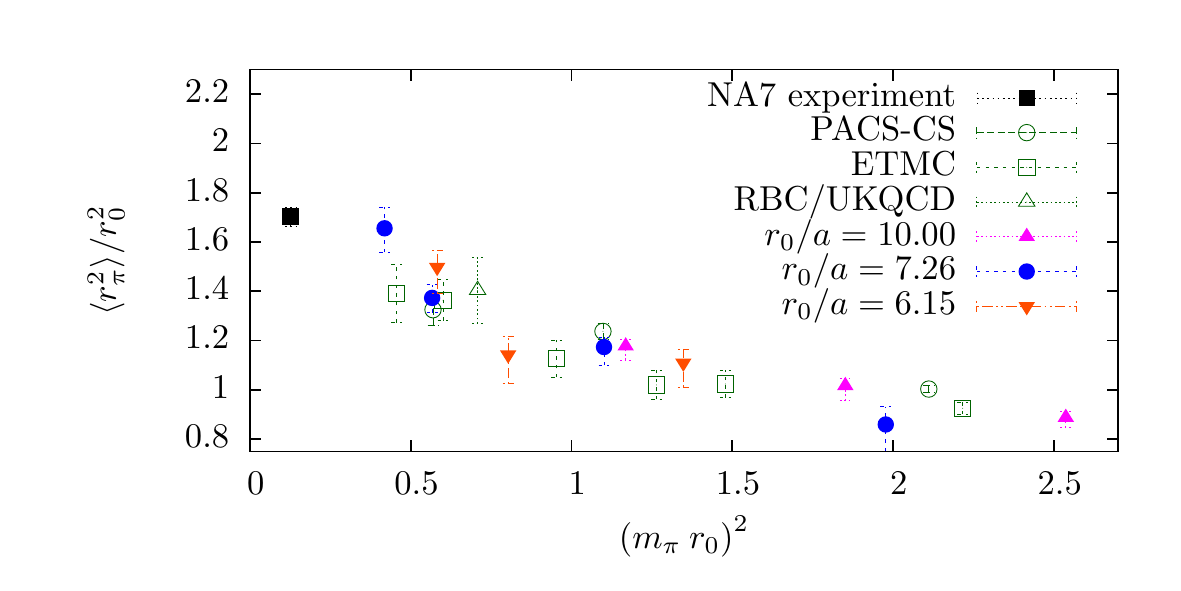}
\end{minipage}
\vspace*{-3mm}
\caption{{\bf Left:} The squared charge radius versus the maximal value of $Q^2$
entering the fit at a pion mass of 295~MeV (ensemble F6). Shown are the results
from a linear fit, polynomial fits of degrees two and three, and the
results of a fit to the vector pole dominance model (VPD). {\bf Right:}
Compilation of results for the charge radius in dynamical lattice
QCD~\cite{Nguyen:2011ek,Frezzotti:2008dr,Boyle:2008yd} and from
experiment~\cite{Amendolia:1986wj}.}
\label{fig:chrad}
\end{figure}

From now on we constrain the data to those kinematical situations where finite
size effects are known in $\chi$PT. To extract the charge radius we compare the
results from linear fits to
the results for the slope at $Q^2=0$ obtained from polynomial fits to order two
and three up to some maximal value of $Q^2$. The results from those different
fits for ensemble F6 are shown in figure~\ref{fig:chrad} (left). We also show
the results obtained from the vector pole dominance model, i.e. of fits to a
single pole form, which has predominantly been used to extract the charge radius
from the data of the form factor. Surprisingly, the linear fit starts to deviate
from the other functional forms at relatively small values of $Q^2$, which shows
that higher order terms become important already around $(Q\:r_0)^2\approx 0.1$.
Since we like to extract the charge radius using a fit to the same $Q^2$
interval for all ensembles we have to use the fit to a polynomial of degree
two with a maximal value of $(Q\:r_0)^2=0.22$, as indicated by the vertical
dashed line in figure~\ref{fig:chrad} (left). In this case all ensembles have
more than three data points for the form factor below this $Q^2$-cut. Note, that
this does not affect the model independent extraction of the charge radius,
since we can explicitly check for the other ensembles that the results are
consistent with the ones from a linear fit to the very small $Q^2$-region. The
results for the charge radius are shown in comparison to results from
experiment~\cite{Amendolia:1986wj} and other
collaborations~\cite{Nguyen:2011ek,Frezzotti:2008dr,Boyle:2008yd} in
figure~\ref{fig:chrad} (right).

\section{Chiral and continuum extrapolation}
\label{sec:chiral}

The remaining task is the extrapolation to the physical point. The natural
framework for the chiral extrapolation is $SU(2)$ $\chi$PT. Since we have found
that $\chi$PT to next-to-leading order (NLO) does not work for our range of pion
masses~\cite{Brandt:2011jk} (for more details see also our upcoming
publication), it is necessary to use $\chi$PT to NNLO. The
expression for the form factor at NNLO in continuum $\chi$PT has been worked out
in~\cite{Bijnens:1998fm}.

For the continuum extrapolation we have to keep track of lattice artefacts,
entering at order $a^2$. To take these into account we include a term
proportional to $Q^2\:a^2$, which respects the normalisation $f_{\pi\pi}(0)=1$.
The resulting formulae then contain 6 free parameters and not all of them can be
constrained sufficiently by a fit to the data for the form factor alone. To
overcome this problem we perform a simultaneous chiral extrapolation of
$f_{\pi\pi}(Q^2)$ (or directly $\left<r_\pi^2\right>$) together with $F_\pi$
and $m_\pi$ with respect to the isospin-averaged light quark mass $\hat{m}$. An
account of the fit formulae will be given in our future
publication.~\footnote{The $\chi$PT part of the formulae is identical with the
ones given in~\cite{Frezzotti:2008dr}.} The inclusion of the additional
quantities further constrains some of the LECs that are less
constrained by the $m_\pi$ and $Q^2$-dependence of the form factor. In total the
formulae including terms modeling cutoff effects now contain 14 free parameters
and further input is needed to be able to obtain reliable results. We therefore
decided to fix the LECs $\bar{\ell}_1=-0.4(5)$ and $\bar{\ell}_2=4.3(1)$ by
$\pi\pi$-scattering~\cite{Colangelo:2001df}. We have verified explicitly that
our fit results are not very sensitive to $\bar{\ell}_1$ and $\bar{\ell}_2$ by
checking that the results for $\left<r_\pi^2\right>$ at the physical point do
not change significantly even if we vary them by a factor two.

To eventually perform the chiral extrapolation we have used three different
procedures, i.e.
\begin{itemize}
 \item simultaneous fits to $\chi$PT to NNLO including $f_{\pi\pi}(Q^2)$,
$F_\pi$ and $m_\pi$,
 \item simultaneous fits to $\chi$PT to NNLO including $\left<r_\pi^2\right>$,
$F_\pi$ and $m_\pi$,
 \item polynomial fits to $\left<r_\pi^2\right>$.
\end{itemize}
The last two procedures make use of the model independent extraction of the
charge radius discussed in the previous section while the first procedure
assumes that $\chi$PT also provides the correct $Q^2$-dependence of the form
factor. Thus the last two procedures are the preferred choices for obtaining a
model independent result for the charge radius at the physical point. The
polynomial fits provide estimates for the systematic effects stemming from the
assumption that $\chi$PT to NNLO provides the correct functional form for the
chiral extrapolation for the present range of pion masses. An estimate for the
associated error can be obtained from the spread of the results of the three
different procedures.

\begin{figure}[t]
\centering
\begin{center}
\vspace*{-5mm}
 \includegraphics[width=0.77\textwidth]{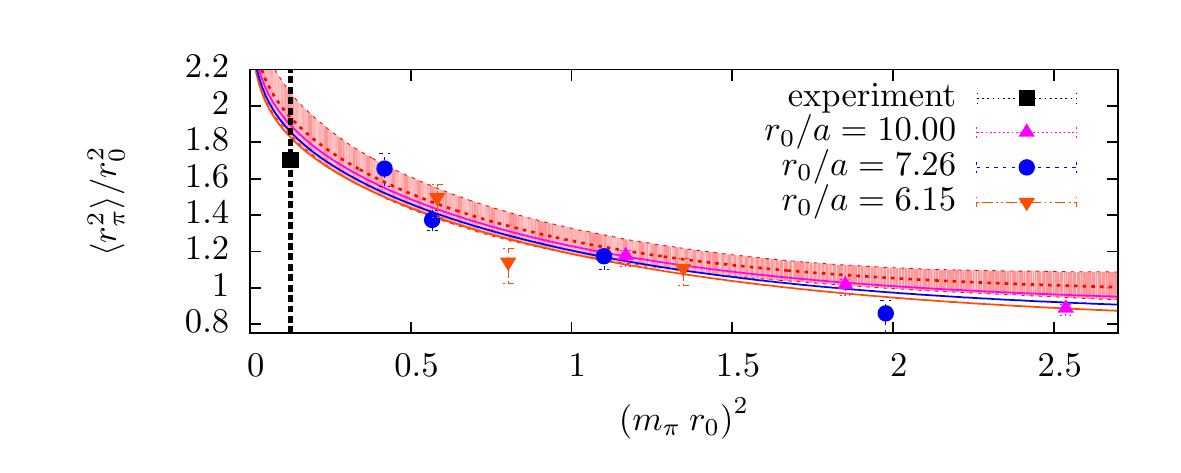} \\[-6mm]
 \includegraphics[width=0.77\textwidth]{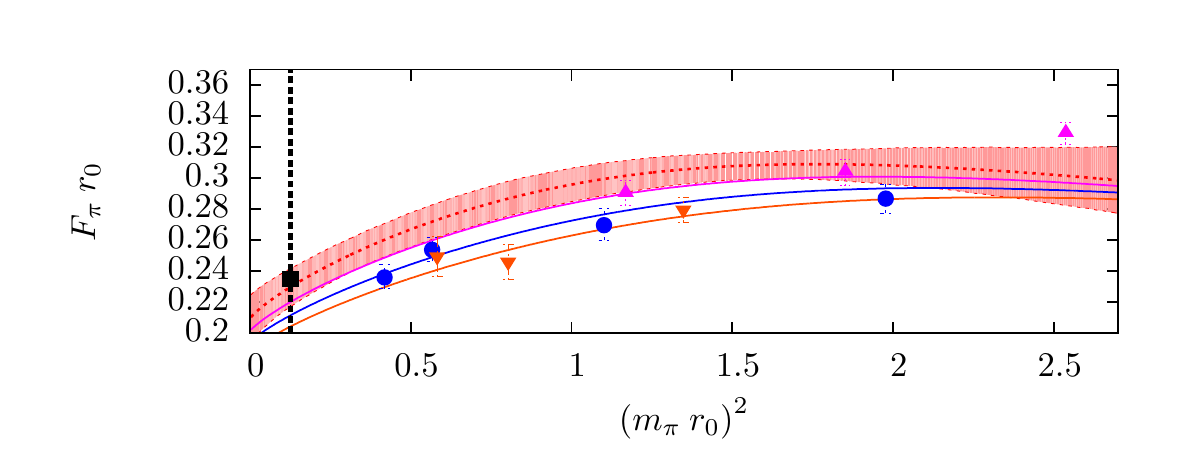} \\[-6mm]
 \includegraphics[width=0.77\textwidth]{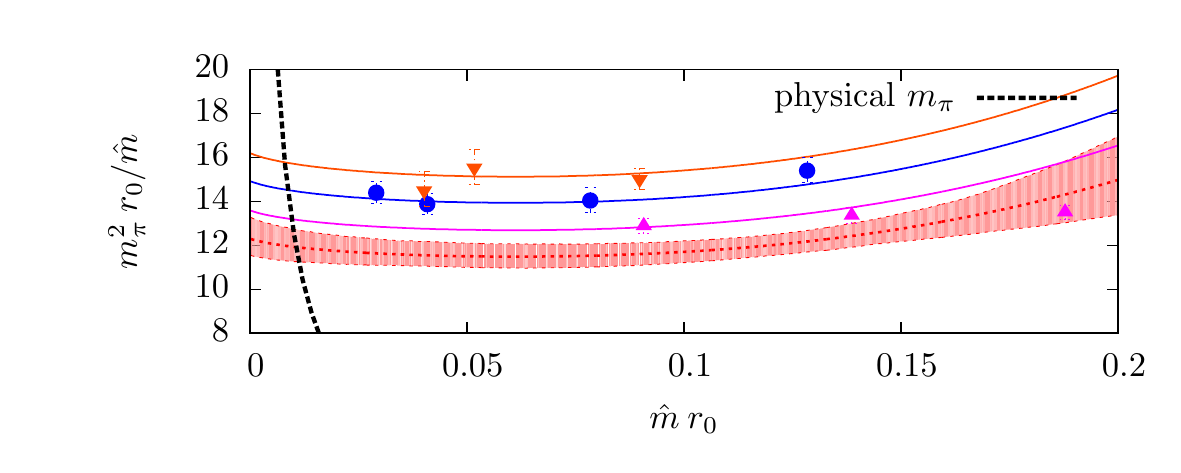}
\vspace*{-4mm}
\caption{Results of a simultaneous fit to the data for $\left<r_\pi^2\right>$,
as extracted in the previous section, $F_\pi$ and $m_\pi$ using $\chi$PT
to NNLO with pion masses below $(m_\pi\:r_0)^2=2$. The coloured dashed lines are
the central values for the different lattice spacings and the red band and the
red dashed line represent the error band and the central value in the
continuum.}
\label{fig:extrapolation}
\end{center}
\end{figure}

Figure~\ref{fig:extrapolation} shows an example for the chiral extrapolation
using $\chi$PT to NNLO for $\left<r_\pi^2\right>$, $F_\pi$ and $m_\pi$ including
only ensembles with a pion mass below 600~MeV. The fit describes the data well
and yields an uncorrelated $\chi^2/\textnormal{dof}=1.3$ with 11 free
parameters. The result in the continuum (red band) and at the physical point
for the charge radius is larger than the experimental value, but not
significantly. The result for the the pion decay constant is consistent with the
experimental result. The same is true for the LECs that are in
good agreement with the typical results found in the literature. A more
comprehensive discussion of the chiral extrapolations will be given in our
future publication.

\section{Conclusions}

In this proceedings article we have given an account of our comprehensive study
of the electromagnetic form factor of the pion at small momentum transfers in
lattice QCD. The high density of accurate data points close to $Q^2=0$ allows
for the model independent extraction of the charge radius at finite lattice
spacings. Finite volume effects have explicitly been studied and corrected using
$\chi$PT to NLO and all quantities are fully $\mathcal{O}(a)$-improved. While
$\chi$PT to NLO fails to describe the data consistently in this range of pion
masses, $\chi$PT to NNLO provides a good description. However, stabilising
the fits is challenging, which agrees with what has been found
in~\cite{Nguyen:2011ek,Frezzotti:2008dr}. Final numbers and detailed
discussions will be given in our upcoming publication.

\acknowledgments
We are grateful to our colleagues within CLS for sharing gauge ensembles. The
calculations of the correlation functions were performed on the dedicated QCD
platform ``Wilson'' at the Institute for Nuclear Physics, University of Mainz.
We thank Dalibor Djukanovic for technical support. The work was supported by
DFG (SFB443) and the Research Center EMG funded by Forschungsinitiative
Rheinland-Pfalz.

\end{document}